\documentstyle[prl,aps,twocolumn,epsfig]{revtex}

\begin{document}
\wideabs{
\title{Laser Cooling of Trapped Fermi Gases deeply below 
the Fermi Temperature}

\author{Z. Idziaszek$^{1,2}$, L. Santos$^1$ and M. Lewenstein$^{1}$}

\address{(1) Institut f\"ur Theoretische Physik, Universit\"at Hannover,
 D-30167 Hannover,Germany\\
(2) Centrum Fizyki Teoretycznej, Polska Akademia Nauk, 02-668 Warsaw, Poland\\
}

\maketitle

\begin{abstract}
We study the collective Raman cooling of a polarized trapped Fermi gas 
in the {\it Festina Lente} regime, when the heating effects associated with 
photon reabsorptions are suppressed. We predict that 
by adjusting the spontaneous Raman emission rates and using 
appropriately designed anharmonic traps, temperatures of the order of 
$2.7\%$ of the Fermi temperature can be achieved in 3D.
 
\end{abstract}

}



The realization of atomic Bose-Einstein condensation (BEC) \cite{BEC}  
and degenerated Fermi gases \cite{Jin,Schreck,Hulet}, 
are among the most remarkable recent achievements 
of the Atomic and Condensed-Matter Physics.
In the context of these experiments it has been recently proposed \cite{Stoof}
that a well-known effect in superconductivity (see e.g. \cite{deGennes}), 
the BCS transition, could be observable in atomic Fermi gases.
The experimental observation of such effect remains, however, 
as an open challenge.
The BCS transition requires temperatures 
well below the Fermi temperature ($T_F$) and,
unfortunately, present techniques, based on evaporative cooling 
in a Fermi gas with two constituents,
have not yet been able to reach the required low temperatures. This is 
due to the Pauli blocking of the 
atom-atom collisions, which prolongs the cooling into time scales at which 
technical losses become dominant \cite{Jin2}. 
Alternative cooling mechanisms \cite{Stringari,Salomon} combined  
with the external modification of the scattering length \cite{Feshbach}, 
have been proposed as possible mechanisms to achieve BCS. 


The aim of this Letter is to show that 
laser cooling offers a realistic and effective method to cool 
fermionic samples well below $T_F$. 
Evaporative cooling \cite{Jin} involves the loss of a large 
fraction of the initial atoms, which in addition lowers $T_F$. 
On the contrary, laser cooling
introduces in principle no atom losses, and therefore allows to cool larger 
fermionic samples, without significantly lowering $T_F$ during the 
process. In addition, laser cooling allows to cool
polarized single--component Fermi gases, 
in which the atom--atom collisions are almost absent. 
Once such a polarized gas was sufficiently cooled, an interaction 
(for instance dipole-dipole one \cite{dipoles}) could be 
externally induced, which can also lead to BCS transition \cite{Misha}.  


Laser cooling of trapped bosons towards BEC have been predicted 
to work in the {\em Festina Lente} (FL) regime \cite{Festina,bosoncool}, 
in which the spontaneous emission $\gamma$ is smaller than the trap 
frequency $\omega$. Experimental confirmation of this possibility has been 
recently reported in Ref.\ \cite{Weiss}. For fermions, apart 
from standard problems (reabsorptions), laser cooling is  
obstacled by the inhibition of spontaneous emission \cite{Busch}. 
Such inhibition, combined with the already slow spontaneous emission rate in 
the FL regime, would lead to unacceptably long cooling times.
This Letter shows that both, reabsorption problem 
and inhibition of the spontaneous emission, can be overcome in the Raman 
cooling of trapped Fermi gases. This is done first by adjusting 
the spontaneous Raman rate, and second by employing the anharmonicity of 
the trap. 


We consider $N$ fermions with an accessible electronic three--level 
$\Lambda$ scheme, with levels $|g\rangle$, $|e\rangle$ and $|r\rangle$. 
We assume that the ground state 
$|g\rangle$ is coupled via a Raman transition to $|e\rangle$ (which 
is assumed metastable). Another laser couples $|e\rangle$ to the upper state 
$|r\rangle$, which rapidly decays into $|g\rangle$. 
After the adiabatic elimination of $|r\rangle$, a two--level system is 
obtained, with an effective Rabi frequency $\Omega$, and an effective 
spontaneous emission rate $\gamma$. The latter can be controlled 
by modifying the coupling from $|e\rangle$ to 
$|r\rangle$. The atoms are confined in a non-isotropic dipole trap with  
frequencies $\omega^g_{x,y,z}$, $\omega^e_{x,y,z}$, different for the 
$|g\rangle$ and $|e\rangle$ states, and non-commensurable one with another. 
The latter assumption
simplifies enormously the dynamics of the spontaneous emission processes
in the FL limit. The cooling process consists of 
sequences of Raman pulses of appropriate frequencies, 
similar to those of Refs. \cite{bosoncool}.  
We assume weak excitation, so that
no significant population in $|e\rangle$ is present. This allows to
adiabatically eliminate $|e\rangle$, and
consequently to consider only the density matrix
$\rho(t)$ describing all atoms in $|g\rangle$, and 
being diagonal in the Fock representation corresponding to the bare 
trap levels.


In principle, an inherent heating is introduced by the reabsorption 
of the spontaneously scattered photons. Fortunately, it has been shown that 
such heating is largely reduced in the FL regime \cite{Festina}.
Due to the Pauli blocking of the $s$-wave scattering, the elastic collisions 
can be considered as absent, as well as the collisional two-body and 
three-body losses. Therefore, the main sources of losses are background 
collisions and photoassociation. For the considered laser intensities and 
atomic densities both loss mechanisms are negligible \cite{footnote}, 
in comparison 
to the losses introduced by the removal of excited--state atoms discussed below.

In this Letter we use the notation of Refs.\ \cite{bosoncool}. 
Using the standard theory of quantum stochastic processes 
one can derive the quantum master equation which 
describes the atom dynamics in the FL regime. 
One can then adiabatically eliminate the excited state, 
to obtain
the rate equations for the populations $N_{j}$ in each level of 
the ground--state trap:
\begin{equation}
\dot N_{n}=\sum_{m}\Gamma_{n\leftarrow m}N_{m}-\sum_{m}\Gamma_{m\leftarrow n}N_{n},
\label{rateeq2}
\end{equation}
where the rates are of the form:
\begin{eqnarray}
\Gamma_{n\leftarrow m}&=&\frac{\Omega^{2}}{2\gamma}\int_{0}^{2\pi}d\phi\int_{0}^{\pi}d\theta{\cal
W}(\theta,\phi) (1-N_{n})\nonumber \\
&\times& \left |\sum_{l}\frac{\gamma\eta_{ln}^{\ast}(\vec
k)\eta_{lm}(k_{L})}{[\delta-(\omega_l^e-\omega_m^g)]+i\gamma 
(R_{ml}+\Delta_{ml})}\right 
|^{2},
\label{Gnm}
\end{eqnarray}
with
\begin{eqnarray}
R_{ml}&=&\int_{0}^{2\pi}d\phi\int_{0}^{\pi}d\theta{\cal W}(\theta,\phi) 
\sum_{n'}|\eta_{ln'}(\vec k)|^{2}(1-N_{n'}),
\label{Rml}
\end{eqnarray}
and $\Delta_{ml}=\int d\phi \int d\theta {\cal W}(\theta,\phi)
|\eta_{lm}(k_{L})|^2$. In the above expressions $2\gamma$ is 
the single--atom spontaneous emission rate,
$\Omega$ is the  Rabi frequency 
 associated with the atom transition and the laser field, 
 $\eta_{lm}(k_{L})=\langle e,l|e^{i\vec k_{L}\cdot\vec r}|g,m\rangle$
 are the Franck--Condon
  factors,  ${\cal W}(\theta,\phi)$ is the fluorescence dipole
  pattern, $\omega _m^g$ 
 ($\omega_l^e$) are the energies of the ground (excited) trap
 level $m $ ($l$), and $\delta$ is 
 the laser detuning from the atomic transition \cite{dip}. 
We consider $\Omega<\gamma$ to ensure the validity of the adiabatic elimination 
of $|e\rangle$.

Note that the rates (\ref{Gnm}) are nonlinear, due to their 
quantum statistical dependence 
on the number of atoms in each level. In particular, we clearly 
distinguish two different quantum--statistical
contributions: (a) 
If the relevant levels are occupied, $R_{ml}$ vanishes, and the atoms remain for very long 
times in the excited state $l$, i.e. the spontaneous emission is inhibited. This 
affects negatively the cooling process in two 
ways: first, the cooling times become very long, and second, the excited-ground collisions 
can occur and lead to heating and losses. In addition, the adiabatic elimination used 
in the derivation of 
Eq. (\ref{rateeq2}) ceases to be valid. 
(b) The fermionic inhibition factor $(1-N_{n})$ appears in 
the numerator of the rates, introducing also a slowing-down of the cooling process.

The above shortcomings can be overcome in the following way. First, note that $\gamma$ 
can be controlled at will in Raman cooling \cite{Festina}. One can 
increase it gradually during the cooling in order to avoid the inhibition effects, 
but remaining in the FL regime.
Still, even if one uses such approach some small fraction of the atoms will remain 
 in the excited state after the cooling pulse, and has to be removed from the trap in order to avoid 
non-elastic collisions. The latter aim can be achieved by optically pumping the 
excited atoms to a third non-trapped level. This introduces a new loss mechanism 
which has to be taken into account in the simulations.

In the following we shall consider a dipole trap 
characterized by a Lamb-Dicke parameter $\eta=2\pi a/\lambda=2$, 
with $a=\sqrt{\hbar/2m\omega}$ being the size of the ground state of the trap, and 
$\lambda$ the laser wavelength. This is the case of current experiments 
in Mg \cite{Ernst}, although our scheme could be also applied to more general traps, 
and other fermionic species.
Let us first consider 1D cooling, as a test of our ideas.
Due to numerical demands in the calculation of the Franck-Condon factors 
for large quantum numbers, we present the results for only $N=200$ 
fermions, $500$ trap levels, and an initial thermal distribution with $T=0.65 T_F$,
We stress, however, that the method could be employed for larger $N$ and an initial $T>T_F$. 
For Mg atoms and a laser wavelength $\lambda \simeq 600$nm, $\eta=2$ 
corresponds to a trap frequency $\omega = 2 \pi \times 5.7$ kHz. 
A sequence of two pulses with respective detunings 
$\delta = -15,-16 \omega$ was applied. When the system becomes degenerated, 
the atoms usually jump after a spontaneous emission from $|e\rangle$ into levels of higher energy, 
since the lower energy levels are occupied.
Therefore large negative detunings of the order of four recoil energies are required. 
Every cooling pulse was assumed to have $\Omega = 0.25\omega$ and a duration  
$t_p = 100 \omega^{-1}$. The spontaneous Raman rate was modified during the 
cooling process in such a way that 
the averaged lifetime of the excited atoms was of the order of $0.05 t_p$. This average was estimated by 
evaluating  the rates (\ref{Gnm}). The atoms which did not return to $|g\rangle$ 
after the cooling pulse were removed from the trap.      

\begin{figure}[ht]
\begin{center}
\epsfxsize=6.0cm
\hspace{0mm}
\psfig{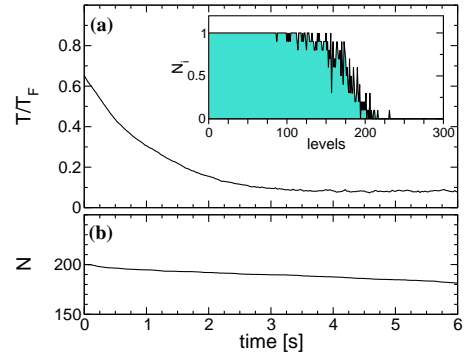}\\[0.1cm]
\caption{1D cooling by using the controlled Raman rate: (a) time dependence of the temperature 
and (b) losses. Inset: final averaged distribution. 
For details, see text.}
\label{fig:1}
\end{center}
\end{figure}

Fig. \ref{fig:1} (a) shows the time dependence of the temperature, whereas in the inset 
the final averaged population distribution is depicted.
Fig. \ref{fig:1} (b) shows the atomic 
losses due to removal of the excited atoms with inhibited spontaneous emission.
As observed, $T\simeq 0.1T_F$ can be reached within 
$4$ s. In order to achieve such low $T$ with relatively small losses 
the value of $\gamma$ has been adjusted from an initial value $0.5\omega$ to $8\omega$ in the final stages.
Note that this does not imply a departure from the FL regime, since the inhibition effects 
drastically reduce the effective width of the excited levels. The temperature was calculated 
from the mean energy $\bar{E}$ with the expression 
$\bar{E}= 0.5 N E_F (1+ (\pi^2/3) (T/T_F)^2)$, valid for low $T$.

To achieve even a lower $T$, 
the excitation of atoms deeply placed in the Fermi sea must be avoided. 
An experimentally feasible method to do it is to use an anharmonic trap, in such a way that 
the Raman lasers are close to resonance just for the atoms within some desired band of energies; only 
those atoms are effectively excited. In particular, for strongly degenerated systems one should only excite 
a band of levels close to the Fermi surface. In the simulations we have used the dependence
$\omega_n^{e,g}=\omega n (1-\alpha n)$ for the trap levels, where $\alpha$ is the anharmonicity parameter 
\cite{foot0}.
Sufficiently small anharmonicity does not affect the Franck-Condon factors, but it does influence the 
resonance term in the denominator of the rates (\ref{Gnm}). 
The requirement that the pulse resonant with the transition $\omega_{m+s}^e-\omega_m^g$, affects only
the atoms occupying the band of energies around $m$, put on $\alpha$ the following constrain: 
$\alpha m_{max}<1/4|s|$ with $m_{max}$ the maximum level considered in the simulation. This inequality 
has to be fulfilled for each cooling pulse.
The width $\delta_m$ of the affected band is related to $\gamma$. Comparison of the two terms 
in the denominator of (\ref{Gnm}) leads to the following formula    
$\gamma \sim \alpha s \delta_m \omega / (4 R_{m,m+s})$ \cite{DiscussR}. 

We have considered a fermionic gas, already pre-cooled by the previous method, with an initial 
$T=0.08 T_F$ and $N=181$ ($10\%$ of the above considered initial number $200$ is lost
during the first stage of cooling). 
The trap parameters were chosen as above, except for the addition of a small  
anharmonicity $\alpha = 1.04 \times 10^{-5}$. Such $\alpha$, and a value 
$\gamma = 0.02\omega$ were chosen to allow for transitions inside a band of about $20$ levels. 
The cooling scheme consist of two pulses with detunings designed to be resonant 
with the transition from the Fermi level ($181$) and $47$ and $48$ levels below respectively. 
Those large detunings are necessary to fill the holes deep inside the Fermi sea by the atoms 
from the surface.
Both pulses have $\Omega = 0.016 \omega$ and a quite long duration $t_p = 5000 \omega^{-1}$. 
The latter is necessary due to the slow Raman spontaneous process in a deeply degenerated system. 
The losses in this
case are negligible ($ \sim 1\%$) mainly due to the long pulses used. 

\begin{figure}[ht]
\begin{center}
\epsfxsize=6.0cm
\hspace{0mm}
\psfig{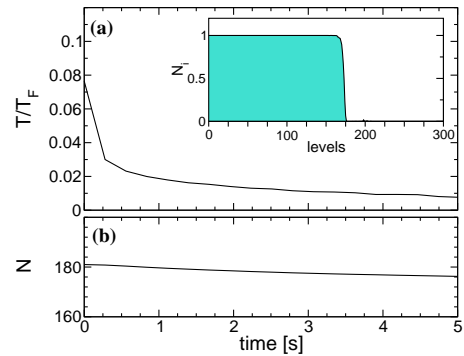}\\[0.1cm]
\caption{
1D cooling by using the anharmonicity of the trap: (a) time dependence of the 
temperature and (b) losses. Inset: final averaged distribution.
For details, see text.}
\label{fig:2}
\end{center}
\end{figure}

Fig. \ref{fig:2} (a) shows the time dependence of the temperature, whereas its inset shows 
the final atom distribution. At the end of this cooling stage (after $5$ s) the temperature 
decreases down to $0.01T_F$. Fig. \ref{fig:2} (b) shows the atomic losses due to the removal 
of the excited atoms with inhibited spontaneous emission.

Finally, we have analyzed the more interesting case 
of a 3D cooling. For simplicity we have considered an isotropic trap, in which the 
level energies have a dependence similar to the 1D case \cite{foot1}. 
We have taken $N=26235$ atoms, $101$ energy shells ($176851$ levels), and an initial
thermal distribution with $T=1.14 T_F$. At $T=0$ the atoms fill $53$ complete energy shells. 
In this case the method of increasing $\gamma$ during the duration of the pulses
was not used, and thus a rather complex sequence
of pulses was employed. The cooling was divided into $7$ stages, 
each one of them consisting in cycles of two Raman pulses with detunings 
$\delta/\omega=\{$$(-6,-7)$, $(-14,-15)$, $(-19,-20)$, $(-19,-20)$, $(-18,-19)$, $(-20,-18)$, $(-20,-18)\}$,
 and tuned to induce transitions from the bands centered at the shells 
$\{16$, $40$, $90$, $70$, $61$, $(53,57)$, $56\}$ with respective widths of
$\{20$, $28$, $60$, $40$, $40$, $(40,10)$, $20\}$ levels.
The spontaneous emission rates were chosen 
$\gamma/\omega=\{$$(0.0075,0.0088)$, $(0.025,0.026)$, $(0.071,0.075)$, $(0.048,0.05)$, 
$(0.045,0.048)$, $(0.48,0.028)$, $(0.15,0.07)\}$.
The detunings were chosen taking into account the 3D 
character of the trap states, since
the excitation quantum number can only be decreased in one direction in one cooling cycle. 
The first four pulses redistribute atoms occupying the levels below $E_F$, making this 
part of the system more degenerate, next six pulses empty the trap levels well above $E_F$ and the 
last four affect the band around  $E_F$. 
We use $\Omega=0.8\gamma$ in each pulse, except in the $11$--th where
$\Omega=0.1\gamma$, to ensure the validity of the adiabatic elimination.

\begin{figure}[ht]
\begin{center}
\epsfxsize=6.0cm
\hspace{0mm}
\psfig{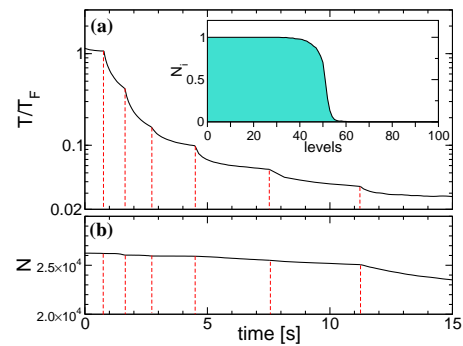}\\[0.1cm]
\caption{
3D cooling by using the anharmonicity of the trap: (a) time dependence of the temperature 
and (b) losses. Dashed lines separate different stages of the cooling process. Inset: 
final averaged distribution. For details, see text.}
\label{fig:3}
\end{center}
\end{figure}

Fig.\ \ref{fig:3} shows: (a) the time dependence of $T/T_F$ during the 
cooling (solid line), and (b) the atom losses.
The temperature was calculated with the help of the grand canonical expressions for the mean values 
of the particle number and energy. At the end of the process ($t=13$s) the 
system reached $T=0.027T_F$. In the simulations we assumed that a fraction of 
the order of $1\%$ of atoms is excited in every pulse, except sixth stage where there was 
about $7\%$ excited atoms in every pulse.
If such fraction is kept constant, similar cooling times are expected for larger $N$, 
the only limitation being the photoassociation losses.

Our analysis can be extended to the case of a multi--component Fermi gas. In such a case, the different 
species 
can interact via $s$-wave scattering, although for a sufficiently degenerated gas such collisions should be 
largely suppressed, and in practice only possible close to the Fermi surface, inducing a 
perturbative modification of the trap levels. 
The collisional dynamics can be split from the laser cooling dynamics \cite{collpaper}, and 
 can be studied within the formalism of Quantum-Boltzmann-Master-Equation \cite{collpaper,Gardiner}. 
The collisions will introduce a thermalization effect, which in fact should help in the cooling.
The case of an unpolarized Fermi gas is particularly interesting when the interaction between 
different species is attractive, since in such a case BCS pairing should be 
possible for temperatures below a critical value $T_c$. 
For reasonable densities and scattering lengths, $T_c$ is
well below $T_F$, and beyond the reach of the present evaporative cooling techniques. The methods discussed 
in our Letter offer an interesting perspective in this sense, since they allow to reach $T\ll T_F$.

Polarized Fermi gases are by themselves also interesting, since once the gas had acquired a 
sufficiently low $T$, an interaction, such as for instance dipole-dipole one
\cite{dipoles}, could be externally induced. Such interaction, partially attractive, allows for the 
appearance of  
BCS transition \cite{Misha}. BCS transition in a dipolar gas 
presents clear advantages compared to the one generated in a two-species gas, 
since for the latter the precise matching of both Fermi surfaces is essential, and experimentally 
very demanding.

Finally, let us point that the full description of the laser-induced BCS transition within 
the Master Equation 
formalism remains as a fascinating challenge. Such analysis will be the subject of future investigations.

We acknowledge  support of the Deutsche Forschungsgemeinschaft (SFB
407), ESF PESC BEC2000+, and TMR ERBXTCT96-0002.
Z. I. was additionally supported by Polish KBN Grant No. 5 P03B 103 20 and the Subsidy of the 
Foundation for Polish Science.
Fruitful discussions with M. Baranov, T. Mehlst\"aubler, J. Keupp, E. Rasel, 
and K. Sengstock are acknowledged.



\begin{references}
\bibitem{dip} Conservative dipole--dipole interactions do not play a 
significant role at the considered densities ($<10^{15}/$cm$^3$), 
and are omitted. Their role is discussed in Refs. \cite{bosoncool,collpaper}.
\bibitem{DiscussR} The coefficient $R_{m,m+s}$ depends on the atom distribution, 
therefore in the calculation of $\gamma$ the effect of increasing degeneracy has to be taken into account.
\bibitem{foot1} 
Anharmonic trap may be designed in such a way 
that the energy levels $\omega_{(n_x,n_y,n_z)}=\omega n (1-\alpha n)$, depend only on 
$n =n_x + n_y +n_z$, i.e. the energy of unperturbed trap.
Such dependence requires the presence of cross terms (like $x^2y^2$) in the perturbative potential.  
Eventually, a potential in which the $x$, $y$ and $z$ dependences 
are separated could also be used, although such potential may lead 
to some accidental degeneracies for transitions well below the Fermi level.
\end{references}
\end{document}